\documentstyle[preprint,tighten,eqsecnum,floats,epsfig,aps]{revtex}

\begin{document}
\draft 

\title{Quantitative complementarity in two--path interferometry}

\author{A. Bramon$^{1}$, G. Garbarino$^{2}$ and B.~C.~Hiesmayr$^{1}$}

\address{$^1$Grup de F{\'\i}sica Te\`orica,
Universitat Aut\`onoma de Barcelona, \\ E--08193 Bellaterra, Spain}

\address{$^2$Departament d'Estructura i Constituents de la Mat\`{e}ria,
Universitat de Barcelona, \\ E--08028 Barcelona, Spain} 
\date{\today}

\date{\today}
\maketitle

\begin{abstract}
The quantitative formulation of Bohr's complementarity  
proposed by Greenberger and Yasin is applied to some physical situations for
which analytical expressions are available. This includes a 
variety of conventional double--slit experiments, but also 
particle oscillations, as in the case of the neutral--kaon
system, and Mott scattering of identical nuclei. 
For all these cases, a unified description
can be achieved  including a  new parameter, $\nu$, which quantifies the  
effective number of fringes one can observe in each specific 
interferometric set--up.   
\end{abstract}

\pacs{PACS numbers: 03.65.-w, 14.40.Aq, 42.25.Hz}

\newpage
\pagestyle{plain}
\baselineskip 16pt
\vskip 48pt

\newpage

\date{\today}
\section{Introduction}

Bohr's complementarity principle and the closely related concept of duality in
interferometric devices are fundamental aspects of quantum mechanics. 
The well known statement that  ``the observation of an interference
pattern and the acquisition of which--way information are mutually exclusive", as
recently rephrased by Englert \cite{englert96}, has been discussed for many years
at a qualitative level.  
Only recently, quantitative statements for this long known
``interferometric duality" \cite{englert96} have become available. 
The first successful step in this direction, which is the one to be considered in the
present paper, is remarkably simple and due to Greenberger and Yasin \cite{GY88}. 
Further extensions and refinements, such as those in the previously mentioned
Ref.~\cite{englert96}, together with experimental analyses, have appeared more
recently and reviewed, for instance, in
Refs.~\cite{englert99,englert00,durr00}.

The quantitative expression for ``interferometric duality" proposed by Greenberger and
Yasin \cite{GY88} reads: 
\begin{equation}
\label{PV0}
{\cal P}^2 + {\cal V}_{0}^2 \le 1, 
\end {equation}
where the equal sign is valid for pure quantum mechanical states. 
In the previous expression, 
${\cal V}_{0}$ is the fringe visibility which quantifies the sharpness 
or contrast of the interference pattern (a wave--like 
property), whereas  ${\cal P}$ is the path ``predictability"   
quantifying the {\it a priori} knowledge one can have on 
the path taken by the interfering system (its complementary 
particle--like property).  
Since we restrict our analysis to two--path interferometers, ${\cal P}$ is defined by 
\cite{GY88}:
\begin{equation}
\label{P}
{\cal P} = |w_{I} - w_{II}| ,
\end {equation}
where $w_{I}$ and $w_{II}$ are the probabilities for taking each path,   
$w_{I} + w_{II} =1$. As already stated, these are {\it a priori} probabilities which depend
exclusively on the state of the interfering system and the specific parameters of the
experimental set--up; in other words, we discard 
information improving measurements as contemplated
in Ref.~\cite{englert96}. The fringe visibility in Eq.~(\ref{PV0}) is defined in 
the standard way. It appears in the oscillatory factor of the intensity, $I(y)$, 
of a generic interference pattern: 
\begin{equation}
\label{I}
I(y)= F(y) \left\{ 1 + {\cal V}_{0}(y) \cos [\phi (y)] \right\} ,
\end {equation}
where $\phi (y)$ is the phase--difference between the two paths, $y$ characterizes the
detector position in the interferometric set--up and $F(y)$ is specific for
each set--up.

In the most simple analyses, ${\cal V}_{0}$ is taken as $y$--independent. But this is
often a too idealized assumption because the probabilities $w_{I}$ and $w_{II}$
for each path generally depend on the detector position. The path predictability
is then $y$--dependent, ${\cal P}(y)$, and so is the fringe visibility, ${\cal V}_{0}(y)$, via
the generalized version of the complementarity relation (\ref{PV0}): 
\begin{equation}
\label{PV}
{\cal P}^2(y)+ {\cal V}_{0}^2(y) \le 1. 
\end {equation} 
The purpose of the present paper is to investigate the physical situations for which 
the expressions for ${\cal P}(y)$,  ${\cal V}_{0}(y)$ and $\phi (y)$ can be 
analytically computed.
As far as we know, this includes interference patterns of various types of
double--slit experiments, the oscillations due to particle mixing shown, among others, 
by the neutral--kaon system and the  
differential cross section for Mott scattering of
identical particles or nuclei. 

These three kinds of phenomena, pertaining to 
distinct branches of physics, and the meaning of the variable $y$, 
linked respectively to a
linear position, a time variable or a scattering angle, are remarkably 
different from each other. 
In spite of this, we can always obtain a unified description in terms of the same
$y$--dependent expressions:  
\begin{equation}
\label{AB} 
{\cal V}_{0}(y) = \frac{1}{\cosh (Ay)} ,\;\; 
{\cal P}(y) = |\tanh (Ay)|,\;\;
\phi (y) = By ,
\end {equation}
where $A$ and $B$ are {\it constants}. In deriving Eqs.~(\ref{AB}) we have assumed that the 
state entering the interferometer is a pure state. As well known, expression (\ref{PV}) is
then  fulfilled with the equal sign, $\cosh^{-2} (Ay) + \tanh^2 (Ay) =1$, for all $y$. 
Mott scattering experiments of nuclei or particles with spin $S \ne 0$ are usually performed
with unpolarized beams described by a density operator proportional to the
identity matrix. In this case, the previous
results are modified to: 
\begin{equation}
\label{y} 
{\cal V}_0(y) = \frac{K}{\cosh (Ay)} ,\;\; 
{\cal P}(y) = 1-K + K |\tanh (Ay)| ,\; \;
\phi (y) = By ,
\end {equation}
with the new constant $K$ ($0 < K <1$) depending on the mixed state.  
The expression (\ref{PV}) is now satisfied as an inequality,  
$1 + 2K(1-K)[|\tanh (Ay)| -1] < 1$, for all $y$ values. 

An interesting consequence of the linear dependence on $y$ of the two arguments  
$Ay$ [in ${\cal V}_{0}(y)$] and $By$ [in the phase $\phi(y)$] is that the oscillatory 
factor in Eq.~(\ref{I}), 
$I(y)/F(y)=1+ {\cal V}_{0}(y)  \cos (By)$, allows for a full characterization of each one of
the cases we consider in terms of a single ratio $R \equiv |A/B|$. This permits 
an easy comparison 
of quite distinct interferometric experiments (see below) and the definition of a new index:
\begin{equation}
\label{nu} 
\nu \equiv 0.264\, \left|\frac{B}{A}\right|=\frac{0.264}{R} ,
\end {equation}
specific for each set--up.
This index has been defined to
estimate the effective number of fringes, i.e., how many of them appear in a given
set--up before the visibility decreases by the usual factor of $e$. 
 
\section{Double--slit experiments}
\label{bart}

\begin{figure}[ht]
\begin{center}  
\includegraphics[width=200pt, keepaspectratio=true]{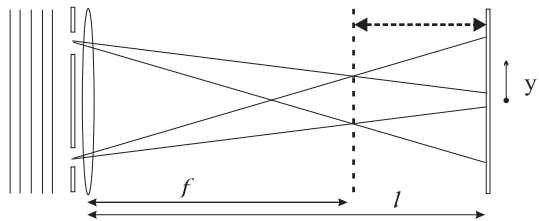}
\vspace{4mm}
\caption{Schematic double--slit set--up considered by Bartell in 
Ref.~\protect\cite{bartell80}
producing the interference pattern of Eq.~\protect(\ref{IDL}). For details,
see the main text.}
\label{bartell}
\end{center}
\end{figure}

\begin{figure}[ht]
\begin{center}
\includegraphics[width=250pt, keepaspectratio=true]{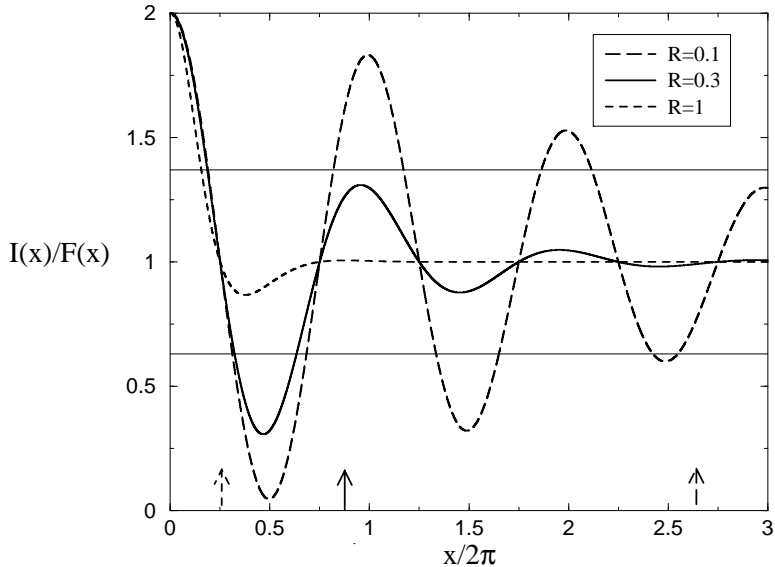}
\vspace{4mm}
\caption{The function $I(x)/F(x)=1+\cos x/\cosh(Rx)$ is plotted
for different values of $R=|A/B|$. The value $R=1$
illustrates the neutral--kaon case of Section~\protect\ref{kaon}. The value
$R=0.3$ refers to Mott scattering of $\alpha$ particles of 
Section~\protect\ref{mott}.
$R=0.1$ corresponds to the double--slit examples of Section~\protect\ref{bart},
as well as to Mott scattering for the nuclei considered in Section~\protect\ref{mott}.
Note that any value of $R$ can be obtained with the double--slit
set--ups by appropriate choices of the experimental parameters.  
Values of $I(x)/F(x)$ between the horizontal lines at $1-1/e$ and $1+1/e$ 
correspond to $|\cos x|/\cosh(Rx)\leq 1/e$.
The vertical arrows correspond to the values of $\nu=0.264/R$ estimating the number 
of observable oscillations and $x$ is in units of $2\pi$.}  
\label{visigen}
\end{center}
\end{figure}

As this is the best known case, a brief analysis should be sufficient for our present
purposes.  We consider the set--up of Fig.~\ref{bartell}, where a monochromatic plane--wave with
wavelength $\lambda = 2\pi /k$ is 
perpendicularly directed towards a double--slit in close contact with a
convergent lens of focal length $f$. The light intensity, 
$I(y)$, is then detected along the $y$--axis of a screen 
perpendicular to the optical axis of the
lens and  placed at a  distance  $l$ from the slits--plus--lens assembly. Assuming a perfect
transparency through the two identical slits (step--function transmission), the analytic
expression for $I(y)$ is well known but it exists only in the Fraunhofer limit, $l=f$, 
where the beams diffracted from each slit overlap and ${\cal V}_{0}=1$. Since we
are interested in the analytic $y$--dependence of ${\cal V}_{0}(y)$, 
we have to discard this simplest
case and follow Bartell's analysis \cite{bartell80}. This requires to consider identical slits
incorporating Gaussian transmission filters, $T(x) = \exp (-x^2 / 2 x_0^2 )$, where $x_0$ is
the effective width of each slit. If their centers are separated by a distance $d$, 
the intensity $I(y)$ along the screen coordinate $y$ is given by 
\cite{bartell80}:
\begin{equation}
\label{IDL} 
I(y) = N e^{- y^2 / \sigma^2} \cosh (Ay) 
\left[ 1 + {1 \over \cosh (Ay)} \cos (By) \right] , 
\end {equation}
where $\sigma^2 \equiv x_0^2 (1 - l/f)^2 + l^2 / (k^2 x_0^2)$   
(the second term here accounts for the spreading of the beam)  
and   
\begin{equation}
\label{ABDS} 
A = {d \over \sigma^2} \left(1-\frac{l}{f}\right) ,\;\;
B = {d \over \sigma^2} {l \over k x_0^2}, \;\; 
R= \frac{k x_0^2}{l} \left(1-\frac{l}{f}\right) .
\end {equation}
As anticipated, $A$ and $B$ are constants and (\ref{PV}) is fulfilled with the equal
sign since we are dealing with pure states. 
For future reference and illustrative purposes, we have plotted the oscillatory factor 
(inside square brackets) of Eq.~(\ref{IDL}) in Fig.~\ref{visigen} 
for the following set of parameters: 
$k = 10^7{\rm m}^{-1}$, $x_0 =10^{-4}$ m, $d= 3 \cdot 10^{-3}$ m, $l =0.1$ m and $f= 0.11$ m. 
These values imply $R \simeq 0.10$ and an effective number of fringes
$\nu \simeq 2.6$, as shown in Fig.~\ref{visigen}. 

\begin{figure}[ht]
\begin{center}
\includegraphics[width=200pt, keepaspectratio=true]{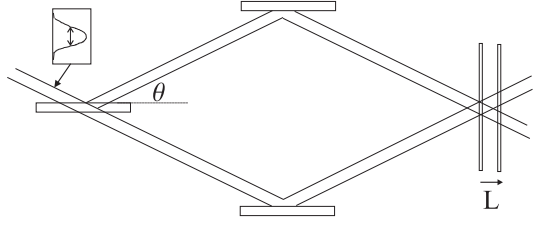}
\vspace{4mm}
\caption{Schematic set--up producing the interference pattern
of Eq.~\protect(\ref{IDL}). A laser beam with Gaussian profile is sent from the
left towards a beam--splitter at a small angle $\theta$.
For details, see the main text.}
\label{bartell2}
\end{center}   
\end{figure}

The need to consider a Gaussian profile for the
beams suggests the use of laser light. 
In Fig.~\ref{bartell2} we have sketched a possible experimental set--up with a symmetrical 
beam--splitter and
two mirrors recombining the beams on a screen. When the latter is at $L=0$, the two
beams completely
overlap and produce interference fringes centered at $y=0$ of maximal visibility 
${\cal V}_0(y) =1$. 
When the screen is displaced ($L \ne 0$) the centers of the two light spots 
separate symmetrically along the $y$--axis and the intensity $I(y)$ is given by 
Eq.~(\ref{IDL}) with $\sigma^2 = x_0^2 $, $A= 2L \sin\theta /x_0^2$, $B = 2k \sin\theta$
and $R= L/ (k x_0^2)$. 
With $k = 10^7 {\rm m}^{-1}$, $x_0 =10^{-4}$ m and $L =0.01$ m one 
obtains again the same $R\simeq 0.10$ and $I(y)$ as before. 

In principle,
neutron interferometric experiments with a double--slit, such as those described in
Ref.~\cite{zeilinger}, could also be considered except for the fact that 
the beam profile is not necessarily
Gaussian and so no analytic expression for $I(y)$ can be derived away from the Fraunhofer
limit. In practice, there is also the problem of correcting our idealized predictions
by the various effects present in the real experiment. As a first attempt, we have
considered that all these corrections can be simulated by a convolution of our 
predictions with an additional Gaussian distribution. Once the single--slit
data of Ref.~\cite{zeilinger} are adjusted in this way, the same convolution
is seen to correct our ideal prediction for the double--slit
and a semiquantitative agreement
with the data of Ref.~\cite{zeilinger} is obtained. 
Quantitative agreement can also be achieved as recently shown in Ref.~\cite{sanz},
but such an analysis goes beyond the scope of the present paper.  Other
neutron interferometric experiments, such as those recently reviewed by
Rauch and Werner \cite{rauch}, could be treated along the same lines once
the corresponding analytical expressions become available. 

\section{Particle oscillations}
\label{kaon}

Particle--antiparticle oscillations are known to take place, among others, in 
the $K^0$--$\bar{K^0}$ system. We restrict our discussion to this case because it has been
considered as the archetypal example \cite{feynman} and studied in detail (for a classical
review, see Kabir's book \cite{kabir}; for recent quantum mechanical results,
see Ref.~\cite{BG02}). 
However, our findings can be immediately extrapolated to 
the $B^0$--$\bar{B^0}$,
$D^0$--$\bar{D^0}$, $\dots$ systems. Neutral kaons are copiously
produced by strangeness conserving strong interactions and so initially appear  either 
as $K^0$'s (strangeness $+1$) or $\bar{K^0}$'s (strangeness $-1$). But time--evolution
in  free space is governed by the $\{K_S,K_L\}$ basis states diagonalizing the weak
Hamiltonian. For these short-- and long--lived  states one has: 
\begin{equation}
\label{SL}
|K_S\rangle = e^{-i\lambda_S t} |K_S\rangle, \;\; 
|K_L\rangle = e^{-i\lambda_L t} |K_L\rangle, 
\end{equation}
where 
$\lambda_{S,L} \equiv  m_{S,L} -(i/2) \Gamma_{S,L}$ and 
$m_{S,L}$  and $\Gamma_{S,L}$ are the kaon masses and decay widths. 

The time--evolution of initial $K^0$ or $\bar{K^0}$ states is then given by:
\begin{eqnarray}
\label{statet}
|K^0\rangle \to |K^0 (t)\rangle &=& {\sqrt{1+|\epsilon |^2} \over \sqrt{2} (1 + \epsilon )} 
\left[e^{-i\lambda_S t} |K_S\rangle + e^{-i\lambda_L t} |K_L\rangle \right] , \\
|\bar{K^0}\rangle \to |\bar{K^0} (t)\rangle &=& 
{\sqrt{1+|\epsilon |^2} \over \sqrt{2} (1 - \epsilon)} 
\left[e^{-i \lambda_S t} |K_S\rangle - e^{-i \lambda_L t} |K_L\rangle \right]. \nonumber
\end{eqnarray}
The CP--violation effects can safely be
neglected ($\epsilon=0$) for our purposes thus obtaining 
$\langle K_S|K_L\rangle =0$ and 
\begin{eqnarray}
\label{statetnS}
|K^0 (t)\rangle &=& {1 \over \sqrt{2}} 
\left[ {1 + e^{-{1\over 2}\Delta \Gamma t} e^{-i \Delta m t}  
\over \sqrt{1 + e^{-\Delta \Gamma t}}} |K^0\rangle +
       {1 - e^{-{1\over 2}\Delta \Gamma t} e^{-i \Delta m t}
\over \sqrt{1 + e^{-\Delta \Gamma t}}} |\bar{K^0}\rangle \right] , \\ 
|\bar{K^0} (t)\rangle &=& {1 \over \sqrt{2}} 
\left[ {1 - e^{-{1\over 2}\Delta \Gamma t} e^{-i \Delta m t}  
\over \sqrt{1 + e^{-\Delta \Gamma t}}} |K^0\rangle +
       {1 + e^{-{1\over 2}\Delta \Gamma t} e^{-i \Delta m t}
\over \sqrt{1 + e^{-\Delta \Gamma t}}} |\bar{K^0}\rangle \right] , \nonumber
\end{eqnarray}
when reverting to the $\{K^0,\bar{K}^0\}$ basis.  
This is necessary to discuss strangeness measurements
projecting into one of these two basis states. In Eqs.~(\ref{statetnS}) we have defined 
$\Delta m = m_L -m_S$, $\Delta \Gamma = \Gamma_L - \Gamma_S$ 
and normalized to undecayed states at time $t$. 

Strangeness oscillations in $t$ of the $|{K^0} (t)\rangle $ and  
$|\bar{K^0} (t)\rangle$ states 
are easily deducible from Eqs.~(\ref{statetnS}). The oscillation phase is given by 
$\phi(t)=\Delta m\, t$  and the time dependent visibility by: 
\begin{equation}
\label{VK}
{\cal V}_{0} (t)={1 \over \cosh \left( {1 \over 2} \Delta \Gamma t\right)}.
\end{equation}
We then recover the structure of Eqs.~(\ref{I}) and (\ref{AB}) with:
\begin{equation}
\label{ABK}
A = {1\over 2}\Delta \Gamma  , \;\; B= \Delta m , \;\; 
R= {|\Delta \Gamma | \over 2 \Delta m} ,
\end {equation}
and from the experiment one has $|\Delta \Gamma|/(2\Delta m)\simeq 1.05$.
These results and the discussion in the previous paragraph clearly show 
the interferometric characteristics of neutral kaon
propagation in free space, where the $K_S$ and $K_L$ propagating components play the role
of the two interferometric paths \cite{BGH03,CPLEARreview}. 
The path--predictability ${\cal P}(t)$ can be computed once one knows that the state has
survived up to time $t$. The larger is $t$, the more probable is the propagation of the 
$K_L$ component ($\Gamma_S \simeq 579 \; \Gamma_L $). One thus gets:
\begin{eqnarray}
\label{PK}
{\cal P} (t) &=& \left| {1 \over 1 + e^{-\Delta \Gamma t}} - 
{1 \over 1 + e^{+\Delta \Gamma t}} \right|
 = \left|\tanh \left( {1 \over 2} \Delta \Gamma t\right)\right|. 
\end {eqnarray} 
Again, one satisfies Eq.~(\ref{PV}), ${\cal P}^2(t) + {\cal V}^2_0 (t) =1$, 
as expected for pure states like $|{K^0} (t)\rangle$ and $|\bar{K^0} (t)\rangle $. 

$B^0$--$\bar{B}^0$ oscillations have also been observed in recent experiments 
leading to compatible values for $\Delta m_B$ (see, for instance, Ref.~\cite{belle}).
The available data are consistent with $\Delta m_B>>\Delta \Gamma_B$, so that 
$R<<1$ and $\nu >>1$.

\section{Mott scattering}
\label{mott}

For energies below the Coulomb barrier, the scattering of two identical nuclei is  
elastic and exclusively due to electrostatic interactions. The differential cross section can
be analytically computed by suitably modifying Rutherford's formula.  This leads to the 
well known Mott's differential cross section:
\begin{equation}
\label{mott1} 
{d \sigma \over d \Omega} = \left( {Z^2e^2 \over 4 E} \right)^2 
\left\{ {1 \over \sin^4 (\theta /2)}  + {1 \over \cos^4 (\theta /2)} + 
C_S\, {2 \over \sin^2 (\theta /2) \cos^2 (\theta /2)} 
\cos [\eta \ln \tan^2 (\theta /2 )] \right\} , 
\end {equation}
where $Ze$ and $S$ are the nuclear charge and spin, whereas $E$ and $\theta$ are the
center--of--mass energy and scattering angle. The first two terms inside brackets
correspond to the squared moduli of the Rutherford scattering amplitudes at angles 
$\theta$ and $\pi - \theta$, respectively. 
The third term comes from the interference of these two
amplitudes and contains the factor $C_S$ accounting for spin effects and the opposite
sign from boson and fermion statistics. 
The two amplitudes can be associated with the direct and crossed
Feynman diagrams, which play the same role as the two paths in conventional double--slit
interferometry. The phase difference appears in the interfering term and depends on the
Sommerfeld parameter:
\begin{equation}
\eta = Z^2 \alpha \sqrt{\frac{Mc^2}{2E}} ,
\end{equation}
$M$ being the mass of the nucleus and $\alpha$ the fine structure constant.

In spite of the obvious differences between this situation and those previously discussed, it is
quite easy to express the $\theta$--dependent differential cross section 
$d \sigma / d \Omega$ in the form of Eq.~(\ref{I}). One needs the change of variable: 
\begin{equation}
\label{cov} 
e^x \equiv {\rm tan}^2(\theta/2)={1 - \cos \theta \over 1 + \cos \theta } ,
\end {equation}
with  $x$ ranging from 0 to $\infty$, as 
$\theta$  runs from $\theta = \pi /2$ to $\theta = \pi $.
Because of the obvious symmetry of
$d \sigma / d \Omega$, negative values of $x$ cover the range 
$0 \le \theta < \pi /2$.  The new variable $x$ allows one
to rewrite the relevant part of Eq.~(\ref{mott1}): 
\begin{equation}
\label{mott2} 
{d \sigma \over d \Omega} 
\propto 1 + {1 - \cos ^2\theta \over 1 + \cos ^2\theta}\; 
C_S \cos [\eta \ln \tan^2 (\theta /2 )] ,
\end {equation}
as:
\begin{equation}
I(x)  \propto 1 + {C_S \over \cosh x}\, \cos (\eta x) ,
\end{equation}
which has the same structure as Eqs.~(\ref{I}) and (\ref{AB}) with:
\begin{equation}
\label{ABn} 
A =1, \;\; B = \eta , \;\; R= 1/ \eta .
\end {equation}
The remaining parameter $C_S$ depends on the
nuclear spin $S$ and is discussed in the following two paragraphs.  

\begin{figure}[ht]
\begin{center}
\includegraphics[width=200pt, keepaspectratio=true]{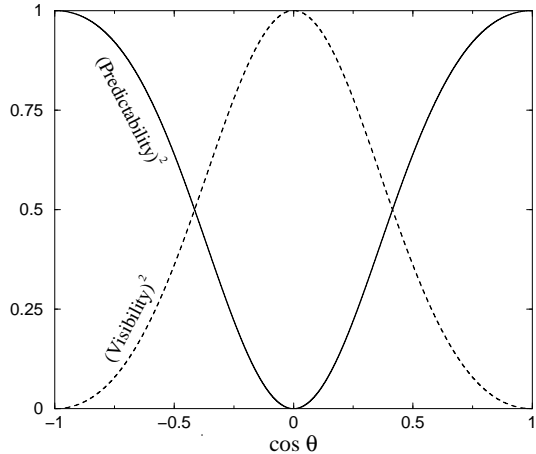}
\vspace{4mm}
\caption{Dependence of the squared values of predictability 
${\cal P}(\theta)= 2|\cos \theta|/(1+\cos^2 \theta)$ 
and visibility ${\cal V}_0(\theta)=(1-\cos^2 \theta)/(1+\cos^2 \theta)$
on $\cos \theta$ for Mott scattering
of spin--zero nuclei or spin--$S$ nuclei polarized in the same direction.}
\label{fig-mott}
\end{center}
\end{figure}

The simplest case to analyze corresponds to the elastic scattering of spin--zero nuclei,
for which one obtains $C_0 =1$ in Eq.~(\ref{mott1}). The same result holds for beams of
spin--$S$ nuclei polarized in the same direction. The equal spin components of 
the two colliding nuclei do not alter their indistinguishability.  
Thus for $\theta = \pi /2$ ($x=0$) the two amplitudes contribute
with the same probability
$w_I =w_{II} = w(\pi /2) = 1/2$, ${\cal P}(x=0)=0$ and one has maximal visibility, 
${\cal V}_0(x=0) = 1$.
For $\theta > \pi /2$ ($x>0$), $w_I = w(\theta )$ and 
$w_{II} = w(\pi - \theta )$ are given  by the Rutherford formula and imply 
${\cal P}(\theta)= 2|\cos \theta|/(1+\cos^2 \theta)$
or ${\cal P}(x)=\tanh x$.  
In Fig.~\ref{fig-mott} we show the dependence
of ${\cal P}^2(\theta)$ and ${\cal V}^2_0(\theta)=(1-\cos^2 \theta)^2/(1+\cos^2 \theta)^2$
on $\cos \theta$.  
These cases admit thus a complete description in terms of pure states
[see Eq.~(\ref{AB})]. 


This is not the case for unpolarized beams of spin--$S$ nuclei. 
The $C_S$ factor appearing in the Mott cross section (\ref{mott1}) is known to be given by  
$ C_S = (-)^{2S}/(2S + 1)$ and the path predictability can be easily computed. 
Indeed, among the $(2S+1)^2$ equally probable spin combinations one can have, 
$(2S+1)$ of them correspond to cases where both nuclei have identical spin components; 
this implies no additional which--path information and the predictability is the same as in
the spinless case. For the remaining $2S(2S+1)$ cases, the paths of either nuclei are 
``marked'' by their distinct spin components and the predictability is one.  
One thus obtains 
${\cal P}(x)= (2S + |\tanh x |)/(2S+1)$ and $K =|C_S|= 1/(2S + 1)$ as required 
[see Eq.~(\ref{y})]. 

\begin{figure}[ht]
\begin{center}
\includegraphics[width=250pt, keepaspectratio=true]{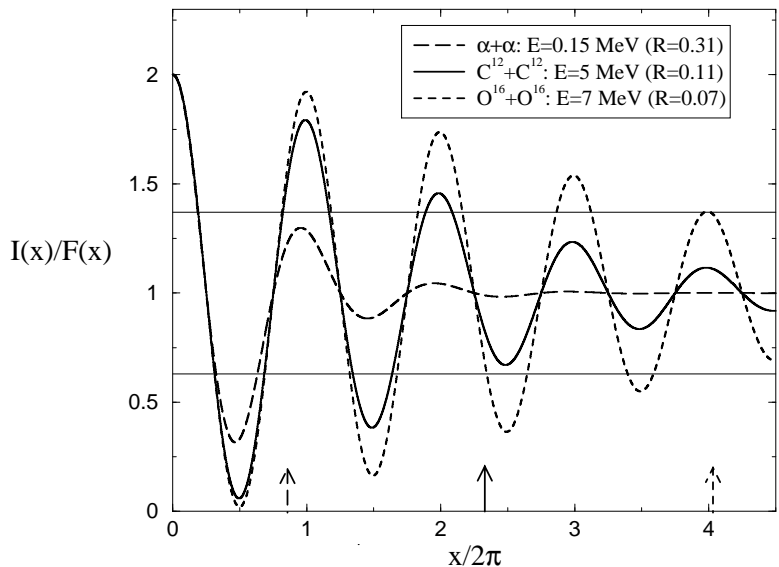}
\vspace{4mm}
\caption{Plot of the function $I(x)/F(x)=1+\cos x/\cosh(Rx)$ for
values of $R=1/\eta$ corresponding to some of the experiments reported in 
Refs.~\protect\cite{HT56,bromley}, where Mott formula was succesfully tested for 
He$^{4}$, C$^{12}$ and O$^{16}$ spin--zero nuclei.
Values of $I(x)/F(x)$ between the horizontal lines at $1-1/e$ and $1+1/e$ correspond 
to $|\cos x|/\cosh(Rx)\leq 1/e$. The vertical arrows correspond to the 
quantity $\nu=0.264/R$ giving the number of observable oscillations.}
\label{visimott}
\end{center}
\end{figure}

Accurate data for Mott scattering below the Coulomb barrier have been obtained in several
experiments long time ago. In Ref.~\cite{HT56}, $\alpha$--$\alpha$ scattering
was measured at center--of--mass energies $E=75$, $150$ and $200$ keV.
Those by Bromley et al. \cite{bromley} refer to ${\rm C}^{12} +
{\rm C}^{12}$ at $E=3$ and $5$ MeV and to
${\rm O}^{16}+{\rm O}^{16}$ at $E=7$, $8.8$ and $10$ MeV. 
In both cases one has $S=0$ and the data are perfectly fitted by the
theoretical curves. The oscillatory factors for some of these curves are 
plotted in Fig.~\ref{visimott}. 
For $S=1/2$ nuclei we have recent data on ${\rm C}^{13} +{\rm C}^{13}$ scattering at 
$E=75$ keV \cite{who?}. These data agree with ${\cal V}_0(x) = 1/ (2\cosh 
x)$ as required by Eq.~(\ref{y}) for $K =1/2$. 
 
\section{Conclusions} 

We have explored the possibility of achieving a unified description for a series of phenomena
belonging to distinct fields of physics but admitting an analogous treatment in terms of basic
concepts of two--path interferometry. This includes the quantitative expression for the
interferometric duality (\ref{PV0}), originally proposed by Greenberger and Yasin, which has
been generalized here to account for the dependence of the fringe visibility 
${\cal V}_0(y)$ and the path
predictability ${\cal P}(y)$ on the detector position $y$. 
We have achieved this unified description for the cases where the
$y$--dependence in ${\cal V}_0(y)$ and 
${\cal P}(y)$, as well as the one in the relative phase $\phi (y)$,
can be analytically computed.
Besides conventional double--slit
phenomena, this includes particle oscillations ---as seen, among others, in the
neutral kaon system--- and the interference effects in the differential cross section of Mott
scattering for identical nuclei. The relevant aspects of the interferometric behaviour of
these various types of phenomena can then be described 
within a global framework and in terms of a unified expression
containing a 
new parameter, $\nu$. This parameter characterizes every specific experimental set--up
and estimates the effective number of visible fringes. 

\section*{Acknowledgements}

This work has been partly supported by EURIDICE
HPRN--CT--2002--00311, MCyT project BFM2002-2588 and INFN.
Enlighting discussions with J.~Campos are acknowledged.



\end{document}